\documentclass[
reprint,
superscriptaddress,
 amsmath,amssymb,
 aps,prl]{revtex4-2}
\usepackage{textcomp}
\usepackage{graphicx}
\usepackage{dcolumn}
\usepackage{bm}
\usepackage{float}
\usepackage{xcolor}
\usepackage{hyperref}
\usepackage{amssymb}
\usepackage{amsmath}
\usepackage{mathrsfs}
\usepackage{hyperref}
\DeclareUnicodeCharacter{2009}{\,}
\begin{document}
\preprint{APS/123-QED}
\title{Diagnosing the impact of relativistically intense prepulse on few-picosecond timeframes for short scale length laser-matter interactions}

\author{H.M. Huddleston}
\affiliation{Centre for Light Matter Interactions, School of Mathematics and Physics, Queen's University Belfast, Belfast BT7 1NN, United Kingdom}

\author{M. Yeung}
\affiliation{Centre for Light Matter Interactions, School of Mathematics and Physics, Queen's University Belfast, Belfast BT7 1NN, United Kingdom}

 \author{C.R.J. Fitzpatrick}
 \affiliation{Centre for Light Matter Interactions, School of Mathematics and Physics, Queen's University Belfast, Belfast BT7 1NN, United Kingdom}
 
 \author{J.P. Kennedy}
 \affiliation{Centre for Light Matter Interactions, School of Mathematics and Physics, Queen's University Belfast, Belfast BT7 1NN, United Kingdom}
 
\author{S. Palaniyppan}
\affiliation{Los Alamos National Laboratory, Los Alamos, New Mexico 87545, USA}

\author{R. Shah}
\affiliation{Los Alamos National Laboratory, Los Alamos, New Mexico 87545, USA}

\author{D. C. Gautier}
\affiliation{Los Alamos National Laboratory, Los Alamos, New Mexico 87545, USA}
\author{M. Zepf}
\affiliation{Helmholtz Institute Jena, 07743 Jena, Germany}
\author{J. C. Fernandez}
\affiliation{Los Alamos National Laboratory, Los Alamos, New Mexico 87545, USA}

\author{B. M. Hegelich}
\affiliation{Center for High Energy Density Science, University of Texas, Austin, TX, 78712, USA}

\author{B Dromey}\email{b.dromey@qub.ac.uk}
\affiliation{Centre for Light Matter Interactions, School of Mathematics and Physics, Queen's University Belfast, Belfast BT7 1NN, United Kingdom}

\date{\today}

\begin{abstract} 
With the rapid proliferation of multi-petawatt (MPW) lasers globally, a new era of high-energy density science promises to emerge within the next decade. However, precise control over how light at these ultra-relativistic intensities interacts with matter (especially with solid-density targets) will be crucial to fully realize the cutting-edge scientific advancements and technological breakthroughs that the MPW regime promises to unlock. In this manuscript, we present experimental results, supported by numerical simulations, which show how intense prepulse activity on few-ps ($10^{-12}$ s) 
timescales leads to rapid shifts in the steepness of the preplasma generated on the surface of ultra-thin nanofoil targets. By combining a single-shot frequency resolved optical gating (FROG) autocorrelation device to diagnose on-shot incident laser pulse contrast, with coherent synchrotron emission (CSE) from relativistic laser plasmas as a probe for evolving plasma-scale length conditions, we provide an experimental benchmark for laser contrast on forthcoming MPW facilities, where high contrast on few-ps timescales will be essential for the next generation of laser-solid interactions.  
\end{abstract}
\maketitle

Multi-petawatt (MPW) laser systems are undergoing rapid international development, aiming to unlock an entirely new era in high-power laser science \cite{danson2019petawatt}. Solid target interactions in the ultra-relativistic regime, including high harmonic generation (HHG), offer potential as a bright coherent source of extreme ultraviolet (XUV) radiation \cite{von1996high, zepf1998role,gordienko2004relativistic,baeva2007high,thaury2007plasma}. Coherent XUV sources are crucial for enabling experiments that explore the development and application of isolated bright attosecond pulses, with key applications in pump-probe studies of real-time fast electron dynamics \cite{hu2006attosecond,tsakiris2006route}. In the case of surface HHG in reflection from solids, the pondermotive force can induce plasma surface denting at relativistic laser intensities. The intrinsic shape of the evolving surface alters the divergence properties of reflected harmonics; a phenomenon known as coherent harmonic focusing \cite{gordienko2005coherent, dromey2009diffraction,horlein2009controlling,vincenti2019achieving}. With a MPW source, this focusing effect could pave the way for experimentally demonstrating vacuum breakdown \cite{schwinger1951gauge}, providing a platform to explore fundamentally new physical regimes \cite{sainte2022quantum,zaim2024light}.

Recently, record intensities of $10^{23}$W/cm$^2$ have already been achieved at the CoRelS 4-PW facility \cite{Yoon:21}. For this system, temporal contrast ratios reaching $10^{-17}$ up to $160$\,ps and $10^{-12}$ up to $2$\,ps before the main pulse with the use of double plasma mirrors has been presented \cite{Choi:20}. Thus, the temporal contrast is no longer limited by pedestals on the nanosecond scale but rather by few-ps contrast. One common cause of poor few-ps contrast seed pulse misalignment with birefringent ti:sapphire crystals that leads to post-pulses with ps delay After additional amplification and compression, these can emerge as pre-pulses or pedestals in the rising edge \cite{didenko2008contrast,khodakovskiy2016degradation}. At laser intensities approaching $10^{12}–10^{14}$\,W/cm$^2$, the laser-solid interaction is already sufficiently intense that a preplasma is formed, which can modify the target structure if this occurs several picoseconds before the peak of the main pulse arrives \cite{murnane1989high,wharton2001effects}. This effect will become more prominent as laser peak power moves toward the MPW regime. Such examples include 50-PW Gekko-EXA \cite{kawanaka2016conceptual}, 100-PW Station of Extreme Light \cite{peng2021overview}, and the 200-PW Exawatt Centre for Extreme Light Studies (XCELS) \cite{kostyukov2023international}.

Precise control of the preplasma gradient is essential for diverse laser-solid interactions—including ion and electron acceleration, and HHG—as success depends on high intrinsic contrast to ensure reproducibility \cite{kahaly2013direct,kaluza2004influence,mckenna2008effects}. The success of such interactions depends on exceptionally high intrinsic temporal contrast to ensure such experiments can be conducted under controlled and repeatable conditions. This work focuses on experimental results of HHG in transmission from ultra-thin foils. Our experimental results demonstrate that a minor degradation in laser contrast on the few-ps level, particularly in the rising edge of the pulse, significantly disrupts experimental conditions. Poor few-ps contrast becomes increasingly problematic as laser peak power scales upward. We discuss how these results act as a benchmark for few-ps contrast on MPW systems, and how the detrimental effects of few-ps contrast is likely to emerge as a key challenge for the field of laser-solid interactions, with prepulse levels projected to approach the relativistic regime.
\begin{figure*}[t]
    \centering
    \includegraphics[width=1.0\linewidth]{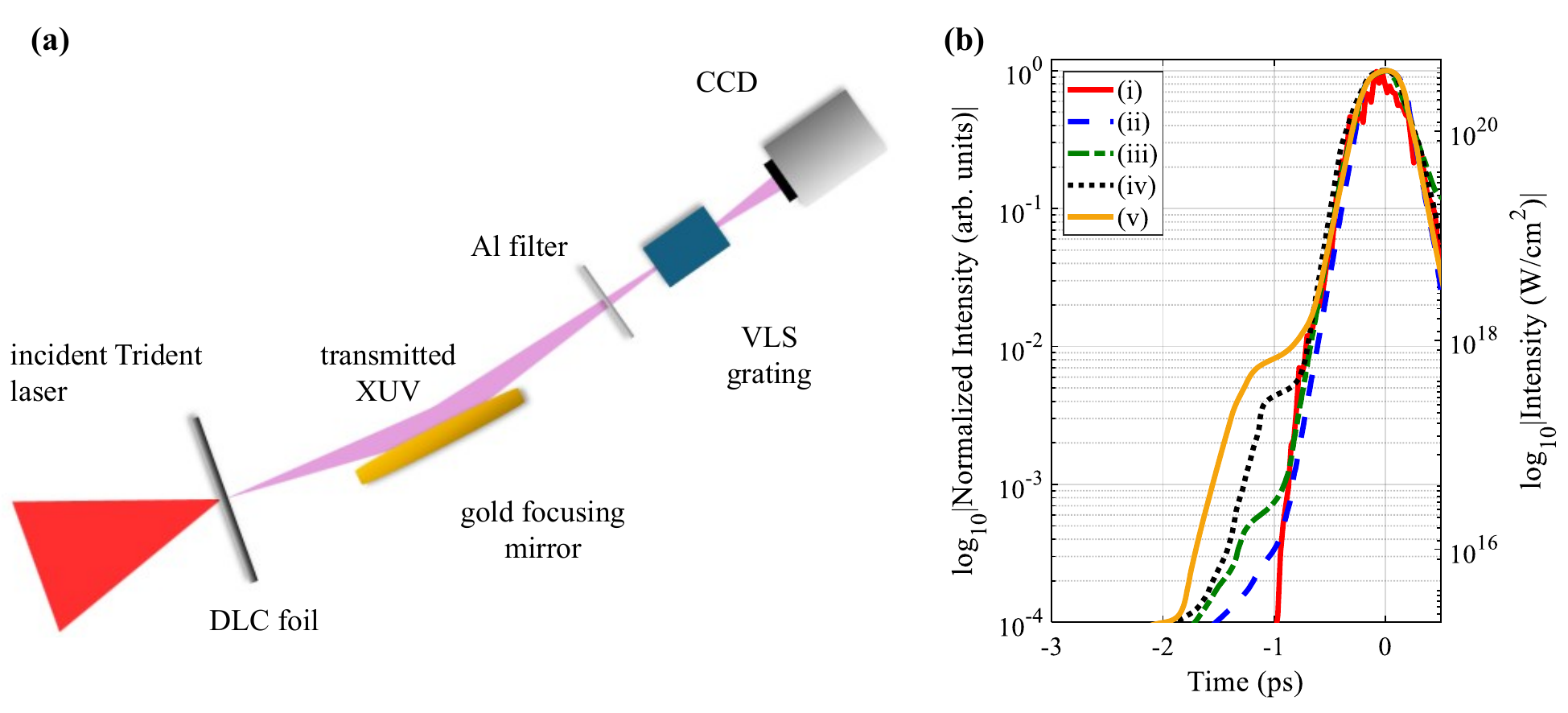}
    \caption{(a) Schematic of the experimental setup. The Trident laser is focused by an \textit{f}/3 off-axis parabolic mirror onto a thin DLC foil. Transmitted radiation is collected by a gold focusing mirror, filtered by a 400\,nm aluminum layer, and detected using a flat-field spectrometer composed of a VLS concave grating and an Andor CCD camera. (b) Shows five temporal traces measured between \(-3.0\,\mathrm{ps}\) and \(+0.5\,\mathrm{ps}\) around the main peak of the laser pulse. These measurements were recorded using single-shot frequency resolved optical gating (FROG) device. The resulting temporal scans gradually reveal a growing pedestal on the rising edge, increasing from shots (i) to (v).}
    \label{fig:1}
 \end{figure*} 
This study focuses primarily on HHG from solid targets. There exist two dominant generation mechanisms observed in reflection for relativistic laser intensity ($a_0^2 = I \lambda_0^2 / 1.37 \times 10^{18} \geq1, \text{W/cm}^{2}\mu\text{m}^2$) where $\lambda_0$ is the central optical wavelength. The first mechanism is a relativistic oscillating mirror (ROM), where the incident pulse induces collective electron oscillations along the plasma surface, periodically driving them to relativistic velocities. The laser pulse is Doppler up-shifted via these oscillations in the over-dense plasma region and reflected specularly \cite{bulanov1994interaction,baeva2007high}. The second mechanism, coherent synchrotron emission (CSE), involves the formation of dense nanometer-scale electron bunches extracted from the plasma by the laser’s electric field \cite{pukhov2010enhanced}. These electron bunches remain tightly compressed and follow synchrotron-like trajectories under the laser field’s influence \cite{cousens2020electron}. The diameter of the electron bunches is considerably smaller than the coherence length for multiple orders of the fundamental wavelength; as such, they emit XUV or synchrotron radiation coherently, up to a wavelength determined by the bunch diameter. 

The preplasma density scale length is a critical factor in observing either mechanism. CSE is more sensitive to preplasma conditions and is normally dominant at much higher intensities than ROM, for which conditions are more relaxed. CSE requires more precise control of the intrinsic laser contrast to achieve and maintain an optimal scale length over a multi-cycle relativistic pulse. Having too soft of a plasma density gradient results in the plasma dynamics evolving during the laser interaction beyond the conditions where CSE harmonics are generated, thus the optimum conditions for CSE are not met with each laser cycle.  

Experimental signatures of both mechanisms are remarkably similar, making it challenging to differentiate between both in reflection. Dromey et al first presented that the CSE mechanism can be isolated using a novel geometry \cite{dromey2012coherent,dromey2013coherent}. Harmonics studied in transmission from a normal incidence interaction with an ultra-thin foil exhibit the shallow spectral scaling consistent with CSE ($I(n) \propto n^{-4/3} $ to $ n^{-6/5}$ where $n$ is the $n$-th harmonic order). Additionally, in this geometry, a sharp cutoff in transmission below the plasma frequency is observed due to the target itself acting as a high-pass spectral filter, thus attosecond pulses are naturally isolated through the interaction. In this letter, we present further observations of HHG in transmission from CSE within this geometry while systematically varying the temporal profile of the laser pulse. This approach allows us to investigate how initial plasma density profile changes influence the resulting spectra.
 \begin{figure*}[t]
    \centering
    \includegraphics[width=1.0\linewidth]{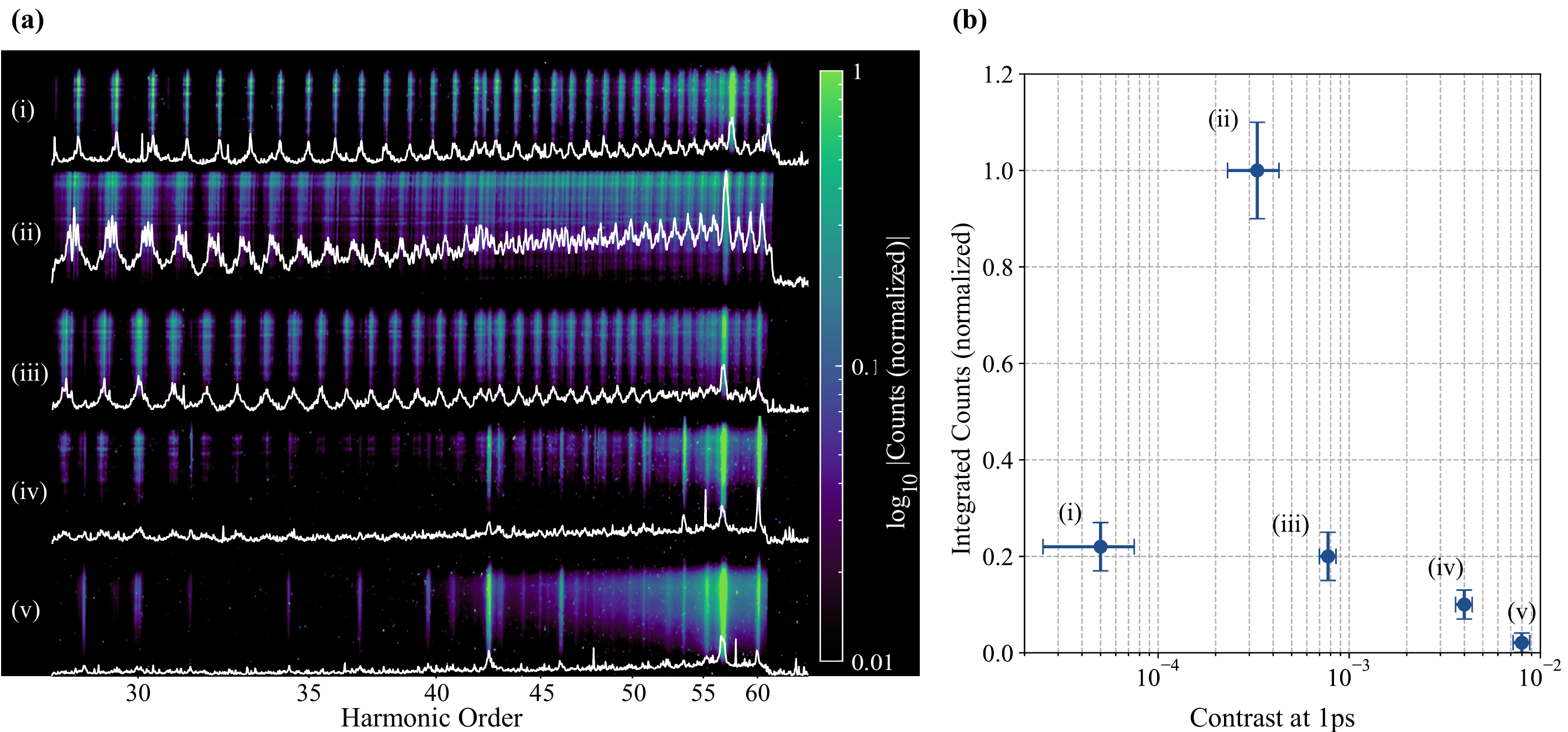} 
    \caption{(a) Harmonic spectra captured by the Andor XUV CCD for each of the five recorded temporal profiles. The white line depicts the harmonic signal averaged across the angular direction (vertical). (b) presents the integrated spectral counts from each CCD image normalized to shot (ii), plotted against the contrast of the pulse -1\,ps before the peak.}
    \label{fig:2}
\end{figure*}
The experiment was conducted at the Trident laser system at Los Alamos National Laboratory \cite{batha2008trident}. The Trident laser, an Nd:glass laser where $\lambda_0 =$ 1054\,nm, delivered 500\,fs pulses with 80 J on target. The experimental setup is illustrated in Fig. \ref{fig:1}(a). The pulses were focused with an $ f/3$ off-axis parabolic mirror, achieving an average peak intensity of around $4 \times 10^{20}$\,W/cm$^2$. 200\,nm thick diamond-like carbon (DLC) foils were irradiated at normal incidence. From the rear surface of the foil, a narrow cone of the transmitted XUV radiation was collected by a gold focusing mirror, passed through a 400\,nm aluminium filter, and detected by a flat-field spectrometer. The spectrometer consisted of a Hitachi variable line spacing (VLS) diffraction grating \cite{kita1983mechanically} and a back-thinned Andor XUV CCD camera.

The Trident laser system exhibited high intrinsic laser contrast low-gain optical parameter amplification (OPA), making it an optimal environment for experiments requiring precise preplasma control without the use of contrast-enhancing plasma mirrors \cite{shah2009high}. Temporal profiles in the few-ps region were measured using a single-shot a single-shot frequency resolved optical gating (FROG) device \cite{palaniyappan2012dynamics}, with temporal resolution of 50\,fs. 

Fig. \ref{fig:1}(b) illustrates the emergence of a pedestal gradually introduced on the order of a ps prior to the peak of the pulse, as captured in the single shot autocorrelation measurements. This contrast degradation varies from $10^{-6}$ to $10^{-2}$ at 1\,ps before the peak. The XUV spectra captured on the CCD from the VLS grating spectrometer for each of these shots are presented in Fig. \ref{fig:2}(a). For clarity, saturated pixels from direct hard X-ray illumination of the CCD have been removed. The normalized integrated signal of the spectra and contrast measured at -1\,ps for each shot is shown in Fig. \ref{fig:2}(b). 

Firstly, focusing on the shot with the sharpest rising edge (i), a spectrum consisting of narrow-band harmonics extends to the aluminum absorption edge after the 61st harmonic (17.1\,nm). This observation is consistent with multiple laser cycles encountering a steep preplasma gradient, maintaining periodic motion of the critical density surface such that a well-defined harmonic comb is produced.

Shot (ii) resulted in an integrated spectrum with total counts almost five times greater than the other shots. A brighter transmitted XUV spectrum is characteristic of a softening of the discontinuous plasma-vacuum boundary, allowing the electromagnetic field to better couple into the plasma particularly on the rising edge of the main pulse. However, focusing on the white line-out of shot (ii) in Fig. \ref{fig:2}(a), the spectrum exhibits spectral broadening of the harmonic peaks, and the real XUV background is significantly raised, indicating that, over the duration of the pulse length, the critical surface motion is changing enough in time to introduce a chirp in the XUV generated each cycle.

Shot (iii) shows a sudden loss in XUV signal, indicating the preplasma gradient has expanded beyond the narrow optimal window. As this few-ps pedestal emerges and begins to cross into the relativistic regime, shots (iv) and (v) illustrate a rapid decay of the harmonic spectrum directly to one typical of a hot plasma generating strong spectral line emission. This rapid transition is a key experimental observation of the sensitive dependence of the CSE mechanism on initial contrast within few-ps time frames.
\begin{figure*}[t]
    \centering
    \includegraphics[width=1.0\linewidth]{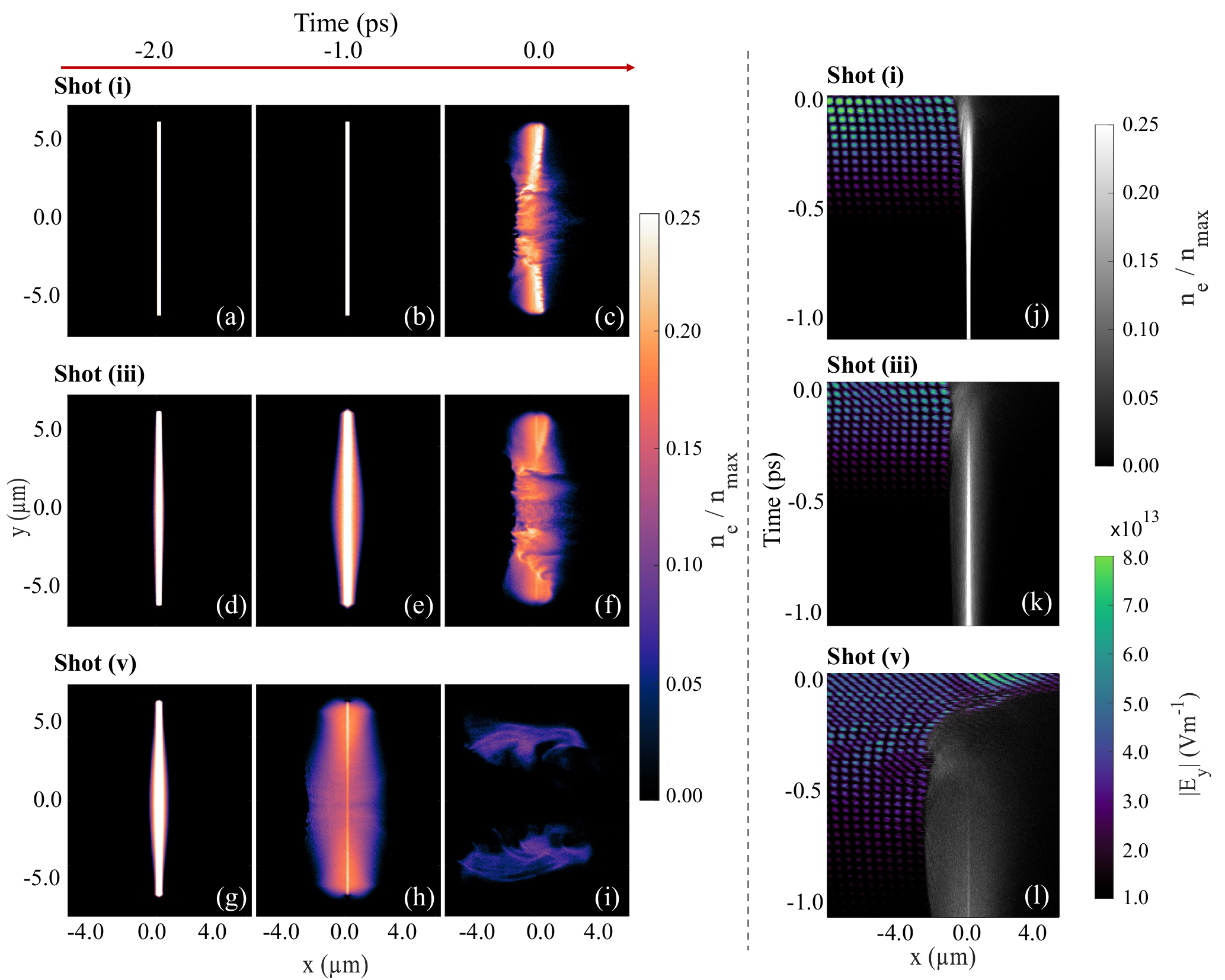} 
    \caption{Snapshots of the electron density of the DLC nanofoil target from 2D particle-in-cell (PIC) simulations, taken in 1\,ps intervals leading up to the arrival of the peak of the pulse $(t = 0\,\mathrm{ps})$. Shots~(i), (iii), and~(v) from Fig. \ref{fig:2} correspond to snapshots~(a--c), (d--f), and~(g--i), respectively. The right-hand plots~(j--l) show the evolution of the electron density along $x$ at the target center $(y = 0)$ over the 1\,ps prior to the laser peak, with the absolute magnitude of the incident electric field overlaid along the axis of propagation as a function of time. As seen in~(l), for a significant pedestal (i.e. shot~(v)), the target expanded to the point where the main pulse is no longer fully reflected. The electron density scales are normalized to the maximum density of the target $(n_{\max} = 800\,n_{\mathrm{crit}})$, where $n_{\mathrm{crit}}$ is the critical density of the plasma. Note that the color scale has been reduced for visibility, so densities above $200n_c$ appear saturated.}
    \label{fig:3}
\end{figure*} 

Large-scale two-dimensional (2D) particle-in-cell (PIC) simulations were performed using EPOCH PIC code \cite{arber2015contemporary} to investigate the effects of the evolving pedestal on the DLC nanofoil targets. Data of the laser pulse shape from around -2\,ps to +0.5\,ps around the peak of the pulse from the experimental data into the simulated pulse incident onto a DLC target with a maximum density of 800\(n_{crit}\), where \(n_{crit}\) is the critical plasma density for Trident's fundamental wavelength. The peak intensities were matched to that of the experiment at  \(4 \times 10^{20}\)\,W/cm$^2$. For higher contrast shots, the intensity of the rising edge began at a level where the collision rate is non-negligible  \((\leq10^{16}\)\,W/cm$^2$) \cite{wilks1993simulations}. Therefore, EPOCH's collision modules were employed \cite{takizuka1977binary, sentoku2008numerical}. A resolution of 4\,nm and 120 particles per cell were implemented to maintain a sufficiently low numerical noise level throughout the simulation.

Fig. \ref{fig:3} presents PIC simulation snapshots of the DLC nanofoil electron density structure and the evolution of preplasma expansion from shots (i), (iii) and (v), which support the degradation of the harmonic spectra observed in the experiment being directly linked to the condition of the preplasma density gradient. Snapshots in 1\,ps intervals up to the peak of the pulse are shown. Figs. \ref{fig:3}(j-l) present the evolution of the electron density at target centre (y=0) during the final picosecond prior to the peak. The electric fields normally incident onto the targets are overlaid to act as a visual aid to observe the consequences of the premature expansion of the DLC targets. The standing wave pattern observed here is a consequence of the interference of the incident and reflected electric field. 

Two main processes causing preplasma expansion at normal incidence are collisional absorption (Inverse Bremsstrahlung) \cite{wilks1997absorption} and relativistic \(\mathbf{J} \times \mathbf{B}\) heating \cite{kruer1985j}. The dominant process strongly correlates to the laser field intensity, as the collision rate becomes negligible as the intensity becomes relativistic. The target with the highest contrast shot (a-c) exhibits a sharp density gradient that is maintained until the main pulse reaches the target. The target remains a reflective surface capable of supporting the electron nanobunching. We see in 3(j) that most of the preplasma formation occurs close to the peak in the rising edge, which is consistent with the experimental harmonic spectrum for this spot. 

In contrast, shot (iii) exhibits a contrast of \(10^{-4}\) at \(-2\) ps. At this intensity \(\approx10^{16}\)W/cm$^2$, collisional absorption is non-negligible, and as the pedestal rises to  \(\approx10^{17}\)W/cm$^2$ at \(-1\) ps, the transition to collisionless $\mathbf{J} \times \mathbf{B}$ heating becomes prominent. A more pronounced preplasma expansion at -1\,ps (Fig. \ref{fig:3}e) with a gradient comparable to the target thickness reflects this transition toward collisionless absorption. Fig. \ref{fig:3}(k), where the rising edge penetrates and interacts with the expanded preplasma, produces a clearly defined high-density surface loss.

Shot (v) maintains a laser contrast of \(10^{-2}\) at \(-1\) ps, placing the pedestal in the relativistic regime where $\mathbf{J} \times \mathbf{B}$ heating dominates. In this case, the preplasma at \(-1\)\,ps (Fig. \ref{fig:3}h) greatly extends beyond the initial target thickness. The decreased plasma density extended across the long gradient in the final picosecond clearly exhibits more rapid heating as the target is completely expanded into a hot plasma in Fig. \ref{fig:3}(i). It is known that for long preplasma gradients, the standing wave patterns from the superposition of the incident and reflected field is interacting with a large under-dense plasma region \cite{chopineau2019identification}. This interaction can trigger a transition from the periodic $2 \omega$ nature of $\mathbf{J} \times \mathbf{B}$ to a more chaotic process like stochastic heating \cite{mendoncca1983threshold,sentoku2002high,kemp2013coupling}. The quasi-static field can couple with the large preplasma volume, contributing to non-periodic XUV emission and the eventual destruction of the intended interaction surface prior to the main pulse arrival. Fig. \ref{fig:3}(l) shows that the bulk of the incident pulse propagates beyond the target in the transmission direction as the plasma expands more rapidly and becomes unstable. The breakthrough of the incident pulse prematurely means that the target will no longer act a free spectral filter for the CSE harmonic beam. 
These results demonstrate the sensitivity of thin targets to subtle shifts in few-ps contrast and underscore that similar constraints apply to a range of ultra-thin target experiments, including radiation pressure acceleration (RPA) where optimal target thickness is on the order of 10's of nanometers.

Recent progress in adapting laser systems for MPW regimes highlights challenges in managing few-ps contrast, especially as systems transition to higher repetition rates. Single-use plasma mirrors enhance pulse contrast post-compression and are designed for preplasma scale lengths below 1\% of the central optical wavelength \cite{dromey2004plasma}. For MPW pulses, however, achieving this requires drastically larger spot sizes to keep prepulses below the ionization threshold. The single-use operation restricts the repetition rate to the refresh rate of the plasma mirror surface, governed by motor speed. Additionally, larger spot sizes would demand significantly larger plasma mirrors or fewer shots per mirror, posing practical limitations for high-repetition-rate operation. Replacing solid plasma mirrors with thin, nanometer-scale thickness liquid crystal films has been reported as a promising development to achieve a high repetition rate. The tunable thickness of the film can also be used to control the contrast on demand \cite{poole2016experiment, zingale2021emittance}.More recently, Edwards et al. proposed a plasma diffraction grating as a high-repetition-rate alternative to plasma mirrors, capable of operating with a switch-on time of approximately 500\,fs and running at up to 10\,Hz \cite{edwards2024greater}. This new technique has not yet been tested or demonstrated for petawatt-class lasers.   

Methods of improving near-time contrast become much more critical for the efficacy of experiments, as with typical facilities currently online reporting contrast  \(>10^{-6}\) a few ps before the peak would result in a relativistic prepulse when scaled to MPW focal intensities. Collisionless absorption effects like $\mathbf{J} \times \mathbf{B}$  and stochastic heating scale will scale with intensity, severely perturbing target conditions. For example, in the context of the experimental data presented in this manuscript, the pedestal of shot(ii) on a 10PW system could become the pedestal of shot(iv), where there is a clear lack of a reflective surface required for a vast range of solid target experiments. Thus, contrast on the few-ps timescales will be required to be up to $10^{10}$ for MPW systems to unlock the yields predicted for short scale length interactions in the ultra-relativistic regime.

In conclusion, we demonstrate that introducing an intense prepulse on few-ps timescales can significantly reduce the efficiency of relativistic laser interaction with ultra-thin foils. We introduce a new diagnostic method for poor few-ps contrast using harmonic spectra from CSE generated by the normal incidence laser interaction with thin carbon nanofoils. We support this via 2D PIC simulations, presenting electron density snapshots of the target in the critical few-ps window before the arrival of the main pulse - which together indicate that even a slight degradation in the contrast of the main pulse on this timescale causes a rapid loss of a sharp narrow-band harmonic spectrum. In the case of relativistic prepulse, collisionless absorption effects such as $\mathbf{J} \times \mathbf{B}$ and stochastic heating can further accelerate preplasma expansion, leading to the destruction of the interaction surface before the main pulse arrives. These results acts as a benchmark for the contrast level required for MPW facilities under development, which will be critically important for unlocking the extreme field science promised in the next generation of laser systems. 
\\
\\
We are grateful for the use of the computing resources from the Northern Ireland High Performance Computing (NI-HPC) service funded by EPSRC (EP/T022175). The EPOCH code used in this work was in part funded by the UK EPSRC Grants No. EP/G054940/1, No. EP/G056803/1, No. EP/G055165/1, No. EP/M022463/1, and No. EP/P02212X/1. M.Y. and B.D. acknowledge support from EPSRC Grant No. EP/W017245/1.
\bibliography{references}

\end{document}